\documentclass{aastex}
\usepackage{emulateapj5,psfig}
\usepackage{graphicx}
\usepackage{timesfonts}
\makeatletter
{\end{center}\end{minipage}\smallskip}

\newenvironment{inlinefigure}{%
\def\@captype{figure}%
\noindent\begin{minipage}{0.999\linewidth}\begin{center}}
{\end{center}\end{minipage}\smallskip}
\makeatother
\def\ltsima{$\; \buildrel < \over \sim \;$}
\def\lsim{\lower.5ex\hbox{\ltsima}}
\def\loe{\lower.5ex\hbox{\ltsima}}
\def\gtsima{$\; \buildrel > \over \sim \;$}
\def\gsim{\lower.5ex\hbox{\gtsima}}
\def\goe{\lower.5ex\hbox{\gtsima}}

\newcommand{\be} {\begin{equation}}

\newcommand{\ee} {\end{equation}}

\newcommand{\VLT}{{\VLT} }

\newcommand{\bc}{\begin{center}}
\newcommand{\ec}{\end{center}}
\def\ltsima{$\; \buildrel < \over \sim \;$}
\def\lsim{\lower.5ex\hbox{\ltsima}}
\def\loe{\lower.5ex\hbox{\ltsima}}
\def\gtsima{$\; \buildrel > \over \sim \;$}
\def\gsim{\lower.5ex\hbox{\gtsima}}
\def\goe{\lower.5ex\hbox{\gtsima}}
\def\sgra{\mbox{SGR~1806--20}}
\def\sgrb{\mbox{SGR~1900+14}}
\def\sgrc{\mbox{SGR~0526--66}}

\newcommand {\rc}{\rm}

\lefthead{ISRAEL ET AL.}
\righthead{RAPID X--RAY OSCILLATIONS IN \sgra}
\submitted{Received: 15 March 2005; ~~ Accepted: 10 May 2005}

\begin{document}

\title{Discovery of Rapid X--ray Oscillations in the Tail of the 
\sgra\ Hyperflare}

\authoremail{gianluca@mporzio.astro.it}

\author{G.L. Israel\altaffilmark{1}, T. Belloni\altaffilmark{2},  L.
Stella\altaffilmark{1},  Y. Rephaeli\altaffilmark{3,4}, D.E.
Gruber\altaffilmark{5}, P. Casella\altaffilmark{2}, S.
Dall'Osso\altaffilmark{1}, N. Rea\altaffilmark{6,1}
M. Persic\altaffilmark{7}, R.E. Rothschild\altaffilmark{4}
}

\affil{1. INAF -- Osservatorio Astronomico di Roma, Via Frascati 
33,   I--00040 Monteporzio Catone (Roma),  Italy, {\small
gianluca@mporzio.astro.it  stella@mporzio.astro.it  dallosso@mporzio.astro.it}}

\affil{2. INAF -- Osservatorio Astronomico di Brera, Via Bianchi 
46, I--23807  Merate (Lc), Italy, \\ {\small belloni@merate.mi.astro.it and
casella@merate.mi.astro.it}}

\affil{3. School of Physics and Astronomy, Tel Aviv University, 
69978 Tel Aviv, Israel, {\small yoelr@wise.tau.ac.il}}

\affil{4. Center for Astrophysics and Space Sciences, University  
of California, San Diego, La Jolla, CA 92093-0424, USA, {\small
rrothschild@ucsd.edu}}

\affil{5. Eureka Scientific Corporation, Oakland CA 94602-3017, USA,  {\small
dgruber@mamacass.ucsd.edu}}

\affil{6. SRON -- National Institute for Space Research, Sorbonnelan 2,
3584 CA, Utrecht, The Netherlands, {\small N.Rea@sron.nl}}

\affil{7. INAF -- Osservatorio Astronomico di Trieste, Via G.B. 
Tiepolo 11, Trieste, Italy, {\small persic@ts.astro.it}}

\begin{abstract}
We have discovered rapid Quasi Periodic Oscillations 
(QPOs) in RXTE/PCA measurements of the pulsating tail of the 27th 
December 2004 giant flare of \sgra.
QPOs  at $\sim$92.5\,Hz are detected in a 50\,s interval starting 
170\,s after the onset of the giant flare. These QPOs appear to be 
associated  with  increased emission by a relatively hard unpulsed component and
are seen  only over phases of the 7.56\,s spin period pulsations away from the
main  peak. QPOs at $\sim$18 and $\sim$30\,Hz are also detected
$\sim$200-300\,s after the onset of the giant flare.
This is the first time that QPOs are unambiguously detected in the flux 
of a Soft Gamma--ray Repeater, or any other magnetar candidate. {\rc We
interpret the highest QPOs in terms of the coupling of toroidal seismic modes
with Alfv\'en waves  propagating along magnetospheric field lines. The
lowest frequency QPO might instead provide indirect evidence on the strength of
the internal magnetic field of the neutron star.}
\end{abstract}


\section{Introduction} 
Soft Gamma-ray Repeaters (SGRs) are characterized by short and recurrent bursts
($<$ 1s) of soft $\gamma$ rays.   Only four 
confirmed SGRs are known, three in the Galaxy and one in the
Large Magellanic Cloud (for a review see e.g. Woods \& Thompson 2004). Giant
flares from SGRs,  with a fluence more than 3 orders of magnitude higher than
that of typical  bursts, are rare. Energy release in the 1979 March 5th event
from \sgrc\ was  $\sim$10$^{44}$ erg (Mazets et al. 1979), while on 1998
August 27th a  giant flare was detected from \sgrb\ at a level of
2$\times$10$^{44}$ erg  (Cline et al. 1998;  Hurley et al. 1999; Feroci et al.
1999). 
The nature of SGRs has remained a mystery for many years. The $\sim 8$\,s
periodicity clearly seen in the tail of the 1979 March 5th giant flare of
\sgrc\  (Mazets et al. 1979)  suggested an association of SGRs  
with  neutron stars. Further evidence came with the discovery and study of the 
persistent counterparts to SGRs in the soft  X--ray range ($<$10 keV). These
display X-ray coherent pulsations at a  period of $\sim 5-8$~s, a secular
spin-down in the range  $\sim$10$^{-11}$--10$^{-10}$ s s$^{-1}$, and a quiescent
X--ray  luminosity orders of magnitude larger than the rotational energy loss.

Several observational properties of SGRs are successfully modelled in 
terms of ``magnetars'' (Duncan \& Thompson 1992; Thompson \& Duncan 1995), 
isolated neutron stars in which the dominant source of free energy is 
their intense magnetic field (B$\sim$10$^{14}$-10$^{15}$ G).
Within this model the short
bursts are  produced by the propagation of Alfv\'en waves in the magnetosphere,
driven  by magnetic field diffusion across small cracks in the neutron star 
crust. The much more energetic giant flares likely arise from a sudden 
reconfiguration of the star's magnetic field that produces large 
fractures in the crust and propagates outwards through Alfv\'en waves of 
enormous power. 
 
On  2004 December 27th a giant flare from \sgra\ was detected by a number 
of  different satellites (INTEGRAL, RHESSI, Swift, Wind, Mars Odyssey, 
etc.; Borkowski et al. 2004, Hurley et al. 2005, Palmer et al. 2005, 
Hurley et al. 2004, Mazets et al. 2004). This ``hyperflare'' caused a 
strong perturbation in the Earth ionosphere and saturated  the detectors on
every high-energy satellite (except  GEOTAIL; Yamazaki et al. 2005, Terasawa et
al.  2005).  A total energy of $\sim 3-10\times$10$^{46}$~erg  (isotropic
emission at a source distance of 8-15 kpc; Cameron et al. 2005;
McClure-Griffiths \& Gaensler 2005) was released  during the main
$\sim$0.2-0.5\,s long spike at the beginning of the event.
Seven days after the event, \sgra\ was observed and detected in the radio 
band for the first time  (VLA; Gaensler et al. 2005a). The radio polarization
and  flux decay was found to be  consistent with synchrotron radiation from an
expanding nebula  (Gaensler et al. 2005b).  

Based on high-time resolution data obtained with the  Rossi X--ray Timing
Explorer (RXTE) Proportional Counter Array (PCA), we report 
the first detailed X--ray timing analysis of the 2004 
December 27th hyperflare of \sgra. In particular, we discover 
fast X--ray Quasi Periodic Oscillations during 
the pulsating tail of the event. Such QPOs are clearly associated with 
a relatively hard emission component which dominates the overall energy 
emission, about 170-220\,s after the beginning of the 
flare.  These QPOs are only detected during a given phase
interval of the 7.56\,s   period, 
\begin{figure*}
\centerline{\psfig{figure=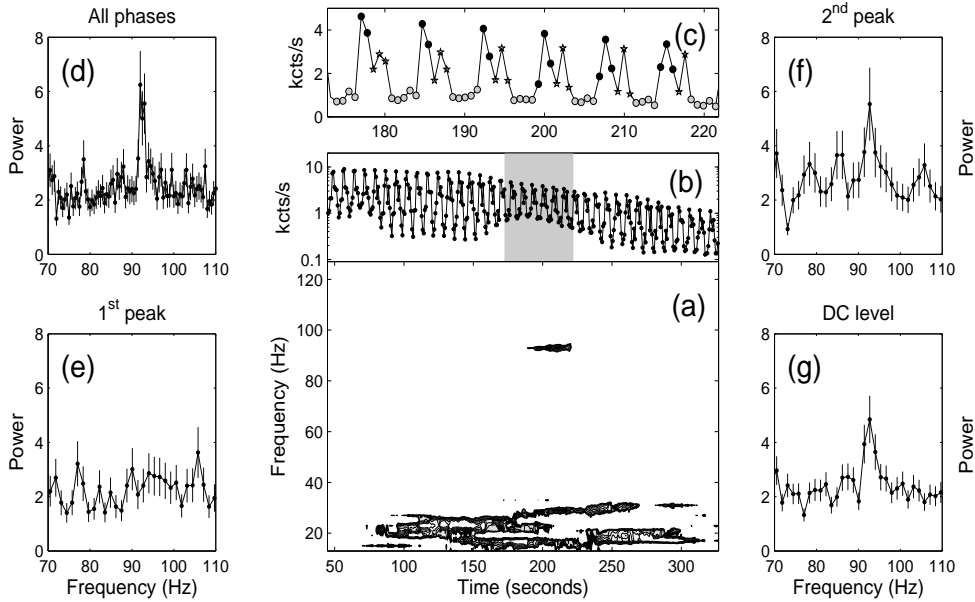,height=9.cm}}
\caption {The $\sim$92.5\,Hz oscillation  in the PCA data. 
(a) Spectrogram with 2\,s time step and resolution. The 
contours represent Leahy powers from 3.2 to 3.7;
(b) Light curve corresponding to the same time axis as panel (a). The
        time resolution is 0.75\,s. The gray-shaded area indicates the
        time interval shown in panel (c);
(c) Close--up of the gray area in panel (b). The different symbols mark
     the  first peak (black circles), the second peak (stars) and the DC 
     level (gray circles);
(d) average power spectrum (at 0.5\,Hz resolution) of the gray area in 
     panel (b); 
(e) average power spectrum (at 1.33\,Hz resolution) of the phase
     interval including the first peak  as seen in panel (c);
(f) same as (e) for the second peak;
(g) same as (e) for the DC level.\label{polittico}}
\end{figure*}

\noindent 
(Fig.\,\ref{polittico}a).  
We briefly discuss some implications of these findings for the magnetar model.

\section{Observations and Timing Analysis}

The giant flare of \sgra\ was recorded by RXTE on 2004 December 27th at
21:31:30.7 UT  during the observation  
of the cluster of galaxies A2163. The PCA was
observing in GoodXenon  mode, resulting in full timing ($\sim 1 \mu s$) and
spectral resolution  (covering the nominal energy range
$\sim$2-80\,keV). Even though the position of \sgra\ was $\sim$30 degrees away
from the RXTE pointing direction  and well outside the field of view, the giant
flare was detected by  the PCA at a peak rate of
$\sim83700$\,counts/s, not corrected for deadtime effects {\rc (note that
photons from \sgra\  propagated through the passive material of the proportional
counter units and collimators before they were recorded by the instruments,
which likely resulted in modification of the incident spectrum)}. The RXTE
telemetry was saturated, gcausing  data gaps in the first $\sim$13\,s of
the flare.  We restricted our timing  analysis to the saturation- and gap-free
interval, starting 12.8\,s after the onset of  the initial spike.  {\rc A search
for anomalous values and/or trends in the housekeeping parameters for the data
in this interval gave negative results.} A light curve was accumulated from all
PCA channels with a resolution of 1/256\,s (a 0.5\,s binned light curve is shown
in Fig. \ref{lc_shape}). Two  spectrograms (also known as
dynamical power spectrum, see Nespoli et al.  2003) consisting of power spectra
built on interval lengths of 2 and 64 seconds, respectively, each shifted from
the other by 2 seconds, were produced. 

A contour plot of the 64\,s  spectrogram is shown in Fig.\,\ref{polittico}a, 
where time 0 is set to T$_0$+12.8\,s, where T$_0$ is the  RHESSI time marking 
the onset of the main peak as  given by Hurley et al. (2005), that is
774026.64 s of TJD 13366. In addition to complex signals below $\sim$30\,Hz
(see below), a peak in the power spectra around 90\,Hz was detected  during the
interval  170-220\,s from T$_0$.
A Lorentzian fit to the peak in Fig.\,\ref{polittico}d 
(limited to the frequency range 70--110\,Hz) yields a centroid frequency of 
92.5$\pm$0.2\,Hz, a FWHM of 1.7$_{-0.4}^{+0.7}$\,Hz (uncertainties are at 
1$\sigma$ confidence level), corresponding to a signal coherence of 
Q$\sim$50. The integrated fractional rms of the peak is 7.3$\pm$0.7\%. The 
single trial chance probability of the 92.5\,Hz QPO peak, calculated by using 
the prescription of Israel \& Stella (1996), is 1.3$\times$10$^{-7}$ which 
reduces to $\sim$1.5$\times$10$^{-5}$ -- 1.3$\times$10$^{-4}$ after 
normalization for the number of periods (128 --  1024) over which the 
search was carried out. We thus consider the 92.5\,Hz QPO detection to be quite
robust. 

In order to investigate a possible dependence of the QPOs on the 
pulsations in the tail of the hyperflare, we adopted a reference period
of 7.5605(6)\,s (Woods et al. 2005) and divided the individual pulsation 
cycles into different phase intervals (see insets in Fig. \ref{lc_shape})
We selected intervals 
centered around the main peak at phase $\sim$0.4, the second peak at phase 
$\sim$0.7, and the minimum around phase 0 (including  
also the third low  peak at phase $\sim$0.15; see insets of Fig \ref{lc_shape}).
The results of 
pulse--phase timing analysis are summarized  in panels  {\it e}, {\it f} and
{\it g} of Fig.\,\ref{polittico}, which show power spectra for the main peak,
the second peak and the minima,  respectively. It is evident that the 92.5\,Hz
oscillations are absent in  the main peak (3$\sigma$ upper limit of
$4.1$\%~rms), clearly detected in  the minima (rms amplitude of $10.7 \pm
1.2$\%), and still present, though at a lower significance, in the second peak
(rms amplitude of $ 8.0 \pm 0.9$\%; no dependence with energy was detected). The
data thus rule out the assumptions that QPO fraction and flux level are
independent of pulsation phase.  The observed pulse phase dependence of the QPO
amplitude makes us confident that the oscillations are intrinsic to
\sgra\footnote{A search for
QPOs  in the light curve of the on-axis source
(A2163) just before and  after the \sgra\ hyperflare also gave negative
results.}.

\begin{inlinefigure}
\psfig{figure=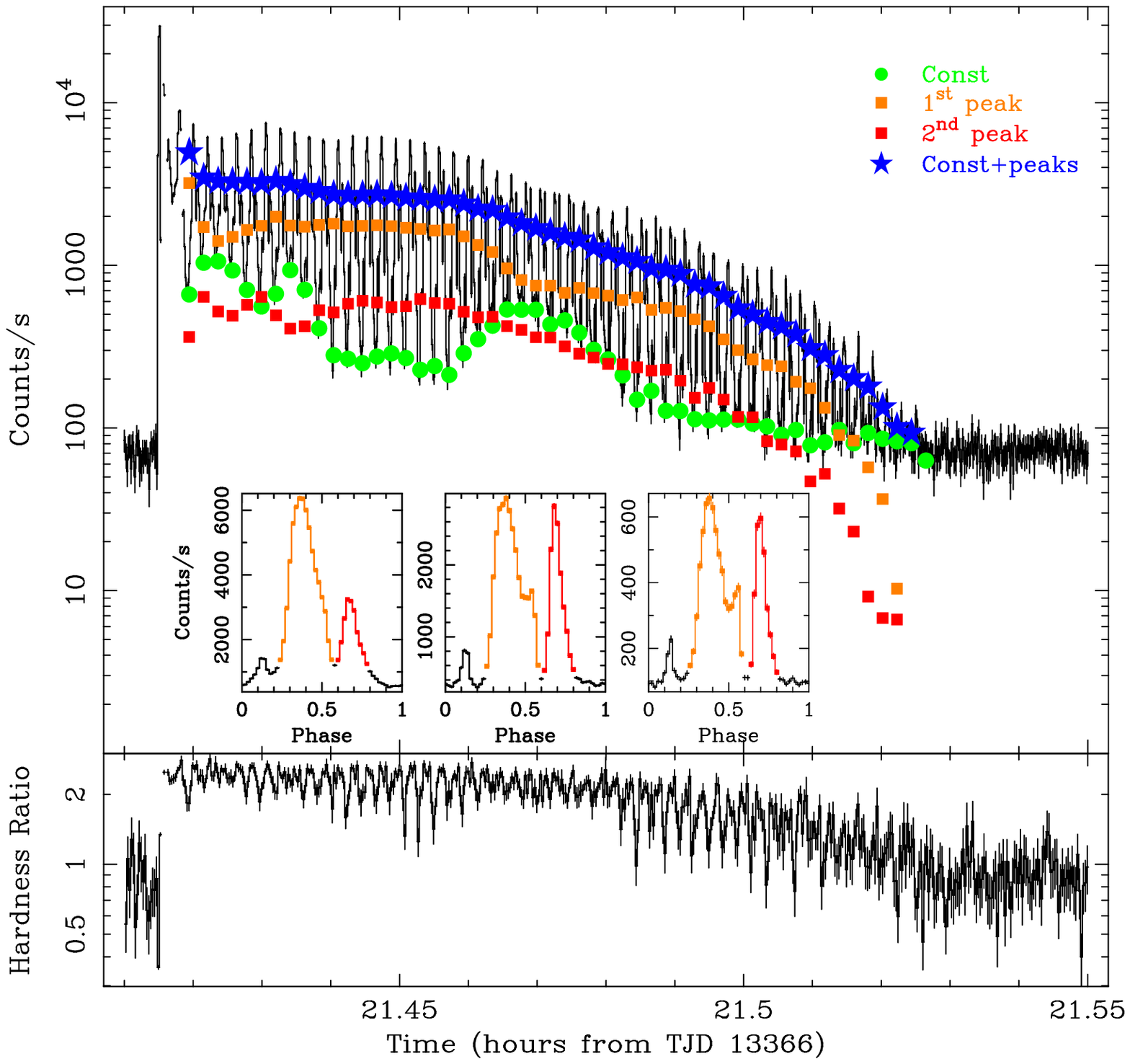,height=8.5cm,angle=-90}
\caption {The RXTE/PCA 0.5\,s binned light curve including the time 
interval of the \sgra\  hyperflare (upper panel). Overimposed are four
parameters  resulting from the fitting of each 7.56\,s pulse with a constant
plus five Gaussian components: the DC level (green circles), the intensity 
of the main peak (orange  squares), of the second peak (red squares), and of all
components (blue stars). The three insets  show the average pulse shapes as a
function of time with the two main peak  marked in orange and red lines,
respectively. Time intervals over which the light  curve have been folded are
approximatively those where insets lie (before,  during and after the gray
region of Fig \ref{polittico}b).  In the lower panel we show  the PCA 9--25\,keV
over 1--9keV energy band light curve ratio (see text for details).
\label{lc_shape}}
\end{inlinefigure}

Since the 92.5\,Hz QPO is detected during the minima of the pulsation cycle
and is absent during the main peak, we checked whether the signal is 
more coherent, but smeared out by {\rc the peak intervals. The
check was  carried out by producing a single power spectrum of the data in the 
gray area of Fig. \ref{polittico}b with pulse peaks replaced with the average
level of the out-of-peak intervals and, therefore,  minimizing the aliasing due
to the pulse oscillation.} The analysis indicates that the 92.5 Hz peak is
indeed intrinsically broad and not consistent with a coherent signal.

From Fig.\,\ref{polittico}a, one can see also a significant amount of power 
in the 20-30\,Hz interval. The structure of this  signal is complex and the
accumulation of all power spectra leads only to a broad excess. We subdivided
the light curve into  smaller intervals  and checked for significant signal
below 40\,Hz. Power spectra obtained from data in the time range
$\sim$200-300\,s (right after  the gray area in Fig.\,\ref{polittico}b), showed
two peaks at  18 and 30\,Hz (see Fig.\,\ref{pds_204_302}). A fit to the
continuum (constant
plus a power--law) yields two QPOs at 18.1$\pm$0.3\,Hz and 30.4$\pm$0.3\,Hz 
with a single trial significance of 3.6 and 4.7$\sigma$, respectively. A 
weak excess is also visible at $\sim$95\,Hz, indicating a possible time 
evolution of the  $\sim$92.5\,Hz QPO frequency. The relatively low statistics in
the time intervals $\sim$200-300\,s prevented us from checking whether these
additional QPOs show a 7.56\,s pulse phase dependence like that displayed by the
92.5\,Hz QPO. 
Finally, we note that the contours in Fig.\,\ref{polittico}a indicate a lower
level of low-frequency power before t$\sim$70. Although a detailed analysis of
the shape of the signal is difficult because of low statistics, it is evident
from the data that before that time the low-frequency noise is steeper and
contributes substantially only below $\sim$10\,Hz.

In the following, we study the 92.5\,Hz QPOs with respect to the other 
parameters of the hyperflare light curve, looking for 
correlations which might provide clues to the origin of the oscillations. To
this end we further analyzed the individual 7.56\,s pulses. Given the highly
non-sinusoidal shape of the pulses we accumulated 50 light curves each
containing one pulse divided into 50 phasebins, and used a multi-Gaussian
fitting model 
produced acceptable fits to all the 50 pulses (we have ignored  the first 2
pulses after the initial spike).  The number of degree of freedom in each fit
was {\rc 33} (for 50 phasebins). Figure\,\ref{lc_shape} shows the result of this
analysis where  the fitted parameters are shown together with the PCA light
curve:  the DC level (green circles), the count rates of the  main peak 
(orange squares), of the second peak (red squares),  
and of all components together (blue stars). From the comparison between 
Fig. \ref{polittico} and Fig. \ref{lc_shape} it is evident that the time 
interval over which the 92.5\,Hz QPOs are significantly detected coincides 
with a bell-shape enhancement in the DC level about 200s after the beginning 
of the flare. Moreover, it is apparent that the intensity of the DC level and 
that of the main peak of the pulse are anti-correlated (i.e., the first peak 
component shows a marked decrease corresponding to the rise of the bell--shaped
intensity bump of the DC), possibly implying two competing  emission
components.  We also note that the  ratio  between the nominal 1-9\,keV
{\rc (effectively, $\sim$8-9\,keV, due to the steep photoelectric  absorption by
the PCA copper collimators)} and 9-25\,keV energy
band light curves shows that the  pulse minima are always softer than the peaks
{\rc (this behavior is similar to that of the \sgrb\ giant flare; Feroci et al.
1999)}, except during the 
bell--shaped bump where minima are nearly as hard as the maxima (see lowest
panel of Fig.\,\ref{lc_shape}). These findings clearly suggest that the 
bell--shaped bump in the decaying pulsating tail of the giant flare 
represents an additional unpulsed component underlying the 
main pulse component.

\section{Discussion}
We discovered rapid quasi periodic X--ray oscillations in the evolving 
X--ray flux of the 2004 Dec 27th hyperflare of \sgra, the first ever for a
magnetar candidate.
The  higher frequency QPOs at  $\sim$92.5\,Hz were detected between 170 and 
220\,s after the onset of the hyperflare, in association with an emission 
bump that occurred in the DC component (and a reduction of the amplitude 
of the 7.56\,s pulsations). These QPOs were detected only in the spin phase 
intervals away from the main peak and reached maximum amplitude corresponding
to  the DC component phase intervals. Evidence for $\sim$18 and $\sim$30\,Hz 
QPOs was found between 200 and  300\,s from the onset of the hyperflare, and not
obviously related to  any specific interval of pulse phases. 

In the context of  the magnetar scenario, the main spike of the giant flare
arises  from a fireball of pair-dominated plasma that expands at relativistic 
speeds, while the energy deposited in the magnetosphere can give rise to a
"trapped fireball" that remains confined to the star's closed magnetic field
lines. The long  pulsating tail of giant flares probably arises from the cooling
of plasma  that remains confined in such a trapped fireball. The bump in the DC
component of the decay some 200\,s after the
main spike  during which the $\sim$92.5\,Hz QPOs were seen, might be due to a
temporary enhancement of the  Alfv\'en wave emission due to dissipation of the
seismic energy. This may lead to the formation of a hot pair-dominated corona
which partially enshrouds  the trapped fireball, thereby reducing the amplitude
of main peak
\begin{inlinefigure}
\psfig{figure=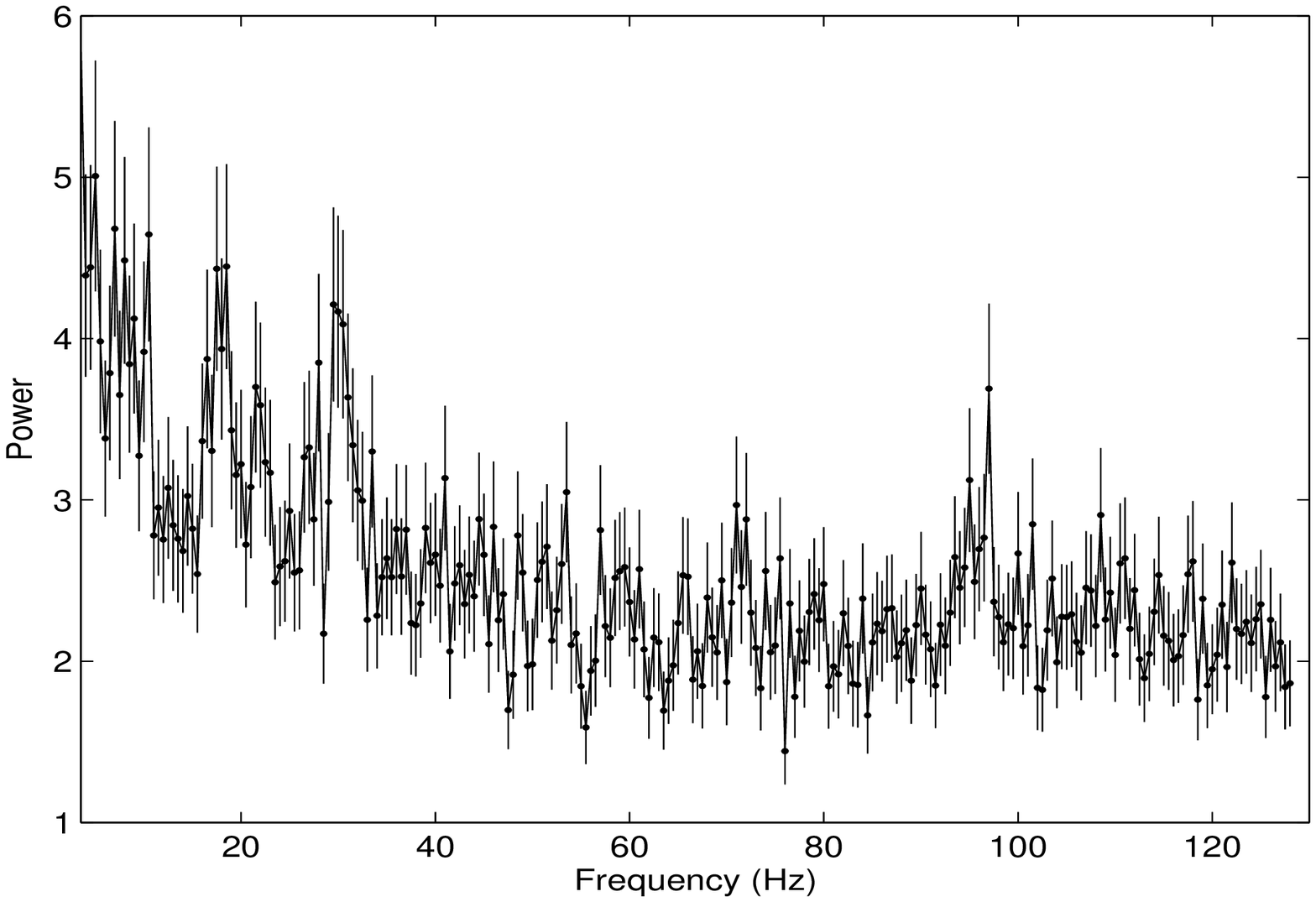,height=6.7cm}
\caption{Power spectrum accumulated from data in the time interval 200--300\,s
(see Fig. \ref{polittico}). Two low--frequency peaks at $\sim$18\,Hz and
$\sim$30\,Hz are visible, together with a small excess at $\sim$95\,Hz (see text
for details).\label{pds_204_302}}
\end{inlinefigure}

\noindent 
 of the 7.56\,s modulation. A similar  interpretation was originally
proposed by Feroci et al. (2001; see also  Thompson \& Duncan 2001) to explain
the smooth flux decay in the first tens of seconds of the 1998 giant flare from
\sgrb. Indeed, a variety  of seismic modes are expected to be excited as
consequence of the  magnetically-induced large scale fracturing of the crust
which gives  rise to giant SGR flares. 
After the main event, the shaking of the neutron star crust by these 
seismic modes gives rise to a coupling with sheared Alfv\'en waves 
on field lines in the neutron star's magnetosphere, in turn causing 
electromotive pair heating (Thompson \& Blaes 1998). 
A modulation of  this heating at the seismic frequencies would easily generate a
flux modulation, as long as the frequencies are sufficiently low that 
phase coherence can  be maintained across the photosphere of the trapped 
fireball. 

Out of the  variety of non-radial neutron star modes studied by McDermott et al 
(1988), there are several classes that have characteristic frequencies 
in the $\sim$10-100\,Hz range. Toroidal modes appear to be especially 
promising because they should be easily excited by the large crustal 
fracturing. Moreover, these torsional  modes couple more easily with the
external magnetic field lines than  modes originating deeper in the stellar
interior (Blaes et al. 1989; Duncan 1998). {\rc  The fundamental toroidal mode
of a rigid neutron star's crust is the $l=2$  ($_2t_0$) mode  whose frequency
corresponds  to a period of $\sim$33.6\,ms, somewhat dependent on the mass,
radius and  crustal magnetic field (McDermott et al. 1988, Duncan 1998). 
The 30.4\,Hz ($\sim$32.8\,ms) oscillation could very well be identified with 
the  $_2t_0$ mode and the 92.5\,Hz ($\sim$10.8\,ms) QPO would thus correspond to
a  higher harmonic: indeed, it matches well the expected frequency of the 
$l$=7 mode, suggesting a relatively small-scale structure in the seismic  wave
pattern (and thus the magnetic multipole structure).  This could be due, for
example, to the principal mode inducing further fractures at various sites in
the crust. 
The shortest duration of the higher frequency QPO is qualitatively in accord
with the expectation that the damping rate of the oscillations strongly
increases with frequency (Duncan 1998). 
Recently,  a large  ($\sim$\,5km) crustal fracturing on the surface of \sgra\
was inferred from a $\sim$5\,ms rise timescale observed during the onset of the
hyperflare (Schwartz et al. 2005). We note that such fracturing can easily
excite the toroidal modes with characteristic frequencies at which QPOs have
been  detected.

The 18\,Hz oscillation, on the other hand, might be associated  with a different
mode which must  couple to the magnetosphere as well. A poloidal component of
the core magnetic field supports a torsional mode with a frequency $\nu_{core}
\simeq 2.5 B_{z,15}$\,Hz (Thompson \& Duncan 2001)  with $R \sim$ 10\,km and a
core density $\simeq  10^{15}$\,g\,cm$^{-3}$, $B_{z,15}$ being the core poloidal
field in units  of $10^{15}$\,G: a strong $B_{z,15} \simeq 7$ would be required
to match the observed 18\,Hz. Although extremely strong, such a field is fully
plausible given that a  (mainly toroidal) field in excess of $10^{16}$\,G is
required to power repeated  giant flares of this magnitude over the $\sim
10^4$\,yr lifetime of an SGR  (Stella, Dall'Osso, \& Israel 2005).}

In summary, the discovery of rapid QPOs in the tail of the 2004 giant 
flare of  \sgra\ {\rc  could constitute the first direct information on the
neutron star crust and magnetic field, thought to be of magnetar strength,}  and
provides a new perspective in the study of neutron star oscillations.

\acknowledgments
This work is partially supported through Agenzia Spaziale Italiana (ASI),
Ministero  dell'Istruzione, Universit\`a e Ricerca Scientifica e Tecnologica
(MIUR -- COFIN), and Istituto Nazionale di Astrofisica (INAF) grants. Work at
UCSD has been supported by a NASA grant. We thank an anonymous referee 
for his/her valuable suggestions.

\end{document}